\documentclass[aps,prl,twocolumn,superscriptaddress,showpacs]{revtex4}
\usepackage{graphicx}
\usepackage{amsmath}
\begin{document}

\title{Radiation pressure acceleration of ion beams driven by circularly polarized laser pulses}

\author{A.~Henig}%
\email{andreas.henig@mpq.mpg.de}%
\affiliation{Max-Planck-Institut
f\"ur Quantenoptik, D-85748 Garching, Germany}%
\affiliation{Department f\"ur Physik,
Ludwig-Maximilians-Universit\"at M\"unchen, D-85748 Garching, Germany}

\author{S. Steinke}
\author{M. Schn\"urer}
\author{T. Sokollik}
\affiliation{Max-Born-Institut, D-12489 Berlin, Germany}

\author{R.~H\"orlein}%
\affiliation{Max-Planck-Institut
f\"ur Quantenoptik, D-85748 Garching, Germany}%
\affiliation{Department f\"ur Physik,
Ludwig-Maximilians-Universit\"at M\"unchen, D-85748 Garching, Germany}

\author{D.~Kiefer}%
\affiliation{Max-Planck-Institut
f\"ur Quantenoptik, D-85748 Garching, Germany}%
\affiliation{Department f\"ur Physik,
Ludwig-Maximilians-Universit\"at M\"unchen, D-85748 Garching, Germany}

\author{D.~Jung}%
\affiliation{Max-Planck-Institut
f\"ur Quantenoptik, D-85748 Garching, Germany}%
\affiliation{Department f\"ur Physik,
Ludwig-Maximilians-Universit\"at M\"unchen, D-85748 Garching, Germany}

\author{J.~Schreiber}%
\affiliation{Max-Planck-Institut
f\"ur Quantenoptik, D-85748 Garching, Germany}%
\affiliation{Department f\"ur Physik,
Ludwig-Maximilians-Universit\"at M\"unchen, D-85748 Garching, Germany}
\affiliation{Plasma Physics Group, Blackett Laboratory, Imperial College London, SW7 2BZ, UK}

\author{B.~M.~Hegelich}%
\affiliation{Department f\"ur Physik,
Ludwig-Maximilians-Universit\"at M\"unchen, D-85748 Garching, Germany}
\affiliation{Los Alamos National Laboratory, Los Alamos, New Mexico 87545, USA}%

\author{X.\,Q. Yan}
\affiliation{Max-Planck-Institut
f\"ur Quantenoptik, D-85748 Garching, Germany}%
\affiliation{State Key Lab of Nuclear Physics and
Technology, Beijing University, 100871, Beijing, China}

\author{T. Tajima}
\affiliation{Department f\"ur Physik,
Ludwig-Maximilians-Universit\"at M\"unchen, D-85748 Garching, Germany}
\affiliation{Photomedical Research Center, JAEA. Kyoto, Japan}

\author{P.\,V. Nickles}
\affiliation{Max-Born-Institut, D-12489 Berlin, Germany}

\author{W. Sandner}
\affiliation{Max-Born-Institut, D-12489 Berlin, Germany}

\author{D.~Habs}%
\affiliation{Max-Planck-Institut
f\"ur Quantenoptik, D-85748 Garching, Germany}%
\affiliation{Department f\"ur Physik,
Ludwig-Maximilians-Universit\"at M\"unchen, D-85748 Garching, Germany}

\date{\today}

\begin{abstract}
We present experimental studies on ion acceleration from ultra-thin diamond-like carbon (DLC) foils irradiated by ultra-high contrast laser pulses of energy 0.7\,J focussed to peak intensities of $5\times10^{19}\,$W/cm$^2$. A reduction in electron heating is observed when the laser polarization is changed from linear to circular, leading to a pronounced peak in the fully ionized carbon spectrum at the optimum foil thickness of 5.3\,nm. Two-dimensional particle-in-cell (PIC) simulations reveal, that those C$^{6+}$ ions are for the first time dominantly accelerated in a phase-stable way by the laser radiation pressure.
\end{abstract}


\maketitle

The generation of highly energetic ion beams from laser-plasma interactions has attracted great interest since the pioneering work that was carried out 10 years ago \cite{ClarkPRL2000, MaksimchukPRL2000, SnavelyPRL2000, HatchettPoP2000}. In the vast majority of previous studies, foil targets ranging in thickness from a few to several tens of microns have been irradiated by linearly polarized, intense ($10^{18}-10^{21}$\,W/cm$^2$) laser pulses. Here, target normal sheath acceleration (TNSA) was found to be the predominant mechanism leading to the emission of multi-MeV, high-quality ion beams. In TNSA, highly energetic electrons are created at the target front surface which traverse the opaque foil to set up a quasi-static electric field of the order TV/m on the nonirradiated side. Protons are preferentially accelerated, resulting in a continuous, exponential spectrum with cut-off energy up to 60\,MeV \cite{RobsonNP2007}. By decreasing the foil thickness to values only slightly above the decay length of the evanescent field, an enhancement in the acceleration of heavier ions was obtained recently \cite{HenigPRL2009b}, being attributed to relativistic transparency at the peak of the pulse, enabling the penetrating laser to transfer energy to all electrons located within the focal volume. Besides raising the maximum ion energy achievable, a second major challenge is imposed by the demand to generate a monochromatic beam. The possibility to accelerate quasi-monoenergetic ion bunches has already been demonstrated within the TNSA-regime by restricting the ion source to a small volume where the sheath field is homogenous \cite{HegelichN2006, SchwoererN2006, Ter-AvetisyanPRL2006}. However, this method suffers from a very low conversion efficiency.\\
Recently, a new mechanism for laser-driven ion acceleration was proposed, where particles gain energy directly from the radiation pressure (RP) exerted onto the target by the laser beam \cite{MacchiPRL2005, ZhangPoP2007, LiseikinaAPL2007, KlimoPRSTAB2008, RobinsonNJP2008, YanPRL2008, RykovanovNJP2008, QiaoPRL2009}, an idea that goes back to \cite{MarxN1966}. For RPA to become dominant, a thin foil is irradiated by a circularly polarized laser pulse at normal incidence. Owing to the absence of an oscillating component in the $\vec{v}\times \vec{B}$ force, electron heating is strongly suppressed. Instead, electrons are compressed to a highly dense electron layer piling up in front of the laser pulse which in turn accelerates ions. By choosing the laser intensity, target thickness and density such that the radiation pressure equals the restoring force given by the charge separation field, the whole focal volume eventually propagates ballistically as a quasi-neutral plasma bunch, continuously gaining energy from the laser field. In this scenario, all particle species are accelerated to the same velocity, which intrinsically results in a monochromatic spectrum. As long as the electron temperature is kept low, a phase-stable acceleration can be maintained, and the process is expected to lead to very high conversion efficiencies and ion maximum energies scaling linearly with laser intensity under optimum conditions.\\
Despite the anticipated highly promising characteristics of RPA, according to our knowledge experimental investigation of ion acceleration using circular polarization has only been carried out at intensities smaller than $4\times 10^{18}$\,W/cm$^2$ and targets of several micron thickness irradiated at oblique incidence \cite{FukumiPoP2005, KadoLPB2006}, thus being far off the parameters necessary for radiation pressure to become the dominant acceleration mechanism.\\
In this Letter, we present for the first time experimental studies of ion acceleration driven by circularly polarized laser pulses at significantly increased intensities of $5\times10^{19}\,$W/cm$^2$. Diamond-like carbon (DLC) foils of thickness $2.9-40\,$nm have been irradiated at normal incidence and ultra-high laser pulse contrast. When compared to the case of linear polarization, we observe a pronounced decrease in the number of hot electrons generated. While for linear polarization the spectra of all ion species decay monotonically up to a certain cut-off value, a distinct peak of energy 30\,MeV emerges in the spectrum of fully ionized carbon ions at the optimum target thickness of $(5.3\,\pm\,1.3)\,$nm and circular polarization. Two-dimensional PIC simulations give evidence, that C$^{6+}$ ions are dominantly accelerated by the laser radiation pressure under these conditions.\\
The described experiments have been carried out at the 30\,TW Ti:sapph laser system located at Max Born Institute, delivering 1.2\,J of energy stored in pulses of 45\,fs FWHM duration at a central wavelength of $\lambda=810\,$nm. The relative intensity of prepulses and of the amplified spontaneous emission (ASE) pedestal was characterized by means of a 3rd order autocorrelator to be smaller than $10^{-7}$ at times earlier than $-10\,$ps prior to the arrival of the main peak. In order to further enhance that value, a re-collimating double plasma mirror \cite{DoumyPRE2004, AndreevPoP2009} was introduced into the laser beam path, resulting in an estimated contrast of $\sim 10^{-11}$. Taking into account the measured $60\%$ energy throughput of this setup, $\sim 0.7\,$J were focussed by an f/2.5 off-axis parabolic mirror to a FWHM diameter focal spot size of $2w_0=3.6\,\mu$m. A peak intensity of $I_0=5\times10^{19}\,$W/cm$^2$ was achieved, corresponding to a normalized laser vector potential maximum of $a_0=5$ for linear and $a_0=5/\sqrt{2}=3.5$ for circular polarization. To vary the laser polarization to circular, a mica crystal operating as $\lambda/4$-waveplate was introduced into the beam path behind the plasma mirror setup.\\
DLC foils of thickness $d=2.9-40\,$nm and density $\rho=2.7$ g/cm$^3$ were placed in the focal plane at normal incidence.
\begin{figure}[t!]
 \includegraphics[width=8.6cm]{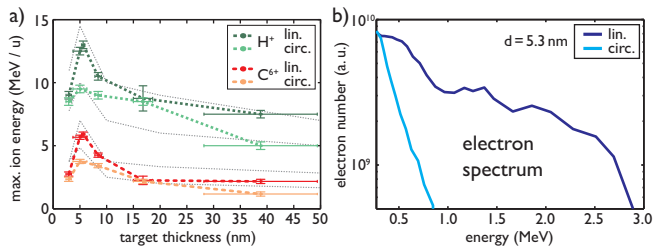}
 \caption{(color). (a) Experimentally observed maximum proton (green, light green) and carbon C$^{6+}$ (red, orange) energies per atomic mass unit over target thickness for linearly and circularly polarized irradiation. Grey dotted lines represent the respective values obtained from 2D PIC simulations. Corresponding electron spectra measured at the optimum target thickness $d=5.3\,$nm are given in (b), showing a strong reduction in electron heating for circularly polarized irradiation. 
 \label{fig1}}
 \end{figure}
Compared to other material available, DLC offers unique properties for mechanically stable, ultra-thin, free standing targets, such as exceptionally high tensile strength, hardness and heat resistance, owing to the high fraction of sp$^3$-, i.e. diamond-like bonds of $\sim$75\%. The thickness of the DLC foils was characterized by means of an atomic force microscope (AFM), including the hydrocarbon contamination layer on the target surface which was present during the experiments. In addition, the depth-dependend composition of the target was measured via Elastic Recoil Detection Analysis (ERDA). From these measurements we obtain a thickness of $\sim 1\,$nm for the hydrocarbon contamination layer. Throughout the manuscript we are referring to the combined thickness of bulk and surface layer as it appears in the actual ion acceleration experiment presented.\\
To characterize the accelerated ions, a Thomson parabola spectrometer was placed at a distance of $\sim0.5\,$m (solid angle $\sim 1.14\times 10^{-7}\,$sr) along the laser propagation, i.e target normal direction. Ion traces were detected at the back of the spectrometer by a CCD camera coupled to a micro-channel plate (MCP) with phosphor screen \cite{Ter-AvetisyanJPDAP2005}. In addition, a magnetic electron spectrometer (solid angle $\sim 2\times 10^{-4}\,$sr) equipped with Fujifilm BAS-TR image plates was positioned behind the target at an angle of 22.5$^\circ$ with respect to the laser axis.\\
The obtained maximum ion energies per atomic mass unit plotted over target thickness $d$ are shown in figure \ref{fig1}a for linear and circular polarization.
\begin{figure}[b!]
 \includegraphics[width=8.6cm]{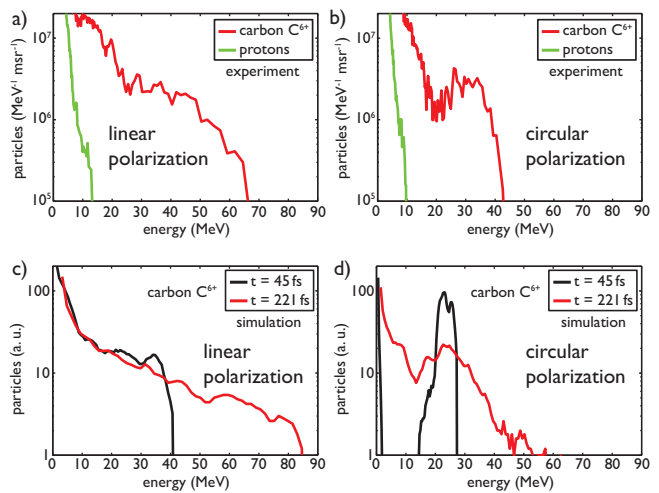}
 \caption{(color). Experimentally observed proton (green) and carbon C$^{6+}$ (red) spectra in case of linear (a) and circular (b) polarized irradiation of a 5.3\,nm thickness DLC foil. The corresponding curves as obtained from 2D PIC-simulations (c,d) show excellent agreement with the measured distributions at late times (red, $t=221\,$fs after the arrival of the laser pulse maximum at the target). A quasi-monoenergetic peak generated by radiation pressure acceleration is revealed for circular polarization, which is still isolated at the end of the laser-target interaction (black, $t=45\,$fs). 
 \label{fig2}}
 \end{figure}
While linearly polarized irradiation yields higher proton and carbon energies, a strong dependence on initial foil thickness is visible in both cases, with a distinct optimum at $d=5.3\,$nm. Using circular polarization, the value of the optimum foil thickness is theoretically expected to be given by the condition $a_0\simeq \sigma$ \cite{YanPRL2008}, i.e., the dimensionless laser vector potential $a_0$ approximately equals the normalized areal density $\sigma=(n_e/n_{cr})\,(d/\lambda)$ of the target. Here, $n_e$ stands for the electron density, whereas $n_{cr}=\epsilon_0 m_e \omega_L^2/e^2$ denotes the critical density of the plasma with electron mass $m_e$ and laser carrier frequency $\omega_L$. This prevision is in excellent agreement with our experimental result of $3.5=a_0\simeq \sigma=3.3$. At the optimum target thickness, maximum energies for protons and carbon ions of 10\,MeV and 45\,MeV are generated for circular polarization, while linear polarization gives 13\,MeV and 71\,MeV, respectively. The corresponding electron spectra for $d=5.3\,$nm are shown in figure \ref{fig1}b. It can be clearly seen, that circularly polarized irradiation results in a pronounced reduction in the number of highly energetic electrons as expected. 
To illustrate the consequent impact on the acceleration of ions, experimentally observed proton and carbon spectra are plotted in figure \ref{fig2}a,b for linear and circular polarization at the optimum foil thickness. A monotonically decaying spectrum is obtained for both protons and carbon ions in case of linear polarization. In contrast, when the laser polarization is changed to circular the spectrum of fully ionized carbon C$^{6+}$ atoms (fig. \ref{fig2}b) reveals two components. In addition to the continuously decreasing low energetic ion population reaching up to $\sim20\,$MeV a distinct peak is seen at higher energies, centered around 30\,MeV. 
This value is nearly identical to the apex energy of 35\,MeV of the quasi-monoenergetic peak of C$^{5+}$ presented in \cite{HegelichN2006}. However, while \cite{HegelichN2006} measured $\sim 1\times10^7$ particles/msr within an energy spread of $\frac{\Delta E}{E}=17\%$ employing a laser pulse of energy 20\,J, we obtain about 50\% more particles within the same spread at a pulse energy of only 0.7\,J, corresponding to a more than 40 times increase in conversion efficiency. 
The spectral spike of C$^{6+}$ ions was repeatedly observed in consecutive shots at the optimum foil thickness of $5.3\,$nm and circular polarization, whereas the shape of the proton spectrum was not affected when varying the polarization.\\
In order to support our experimental findings, two-dimensional particle-in-cell (PIC) simulations were carried out. 
\begin{figure}[t!]
 \includegraphics[width=8.6cm]{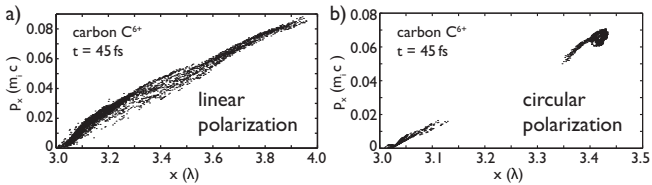}
 \caption{Carbon ion phase space at the end of the laser-target interaction ($t=45\,$fs). A significant amount of particles forms a distinct loop in case of circular polarization (b), giving evidence of a phase-stable acceleration driven by the laser radiation pressure.
 \label{fig3}}
 \end{figure}
The DLC foil targets were modeled by a solid density ($n_e/n_{cr}=500$), rectangularly shaped plasma slab of zero initial electron temperature, composed of 90\% C$^{6+}$ ions and 10\% protons in number density. The laser pulse is of Gaussian shape in both the spatial distribution in the focal plane as well as in time, with a FWHM diameter of $4\,\mu$m and a FWHM duration of 45\,fs, resulting in a peak intensity of $I_0=5\times 10^{19}\,$W/cm$^2$. A simulation domain of size $10\mu$m in transverse ($y$) and $20\mu$m in longitudinal ($x$) dimension was used, subdivided into a grid of $1200\times 10000$ cells each occupied by 2000 particles. In the following, PIC simulation times are given relative to $t=0$, when the peak of the laser pulse reaches the initial position of the target ($x=3\lambda$).\\
The calculated carbon spectra are presented in figure \ref{fig2}c,d for linearly and circularly polarized irradiation. To account for the small solid angle of observation of the Thomson parabola spectrometer, only particles propagating in forward direction within a cone of half angle 0.01\,rad were considered for the simulated graphs. A continuously decaying spectrum is generated in case of linear polarization, which agrees well with the experimental result (fig. \ref{fig2}a,c). This scenario changes drastically when circular polarization is used. As shown in figure \ref{fig2}d, an isolated, quasi-monoenergetic peak emerges in this configuration at the end of the laser-target interaction (black, $t=45\,$fs). In the carbon ion phase space (fig. \ref{fig3}b), a significant amount of particles is located in a discrete area, constituting a rotating structure. The series of loops originates from the continuing front side acceleration and the ballistic evolution of the target, thus giving clear evidence of radiation pressure to be the dominant acceleration force \cite{MacchiPRL2005, ZhangPoP2007, LiseikinaAPL2007, KlimoPRSTAB2008, RobinsonNJP2008, YanPRL2008, RykovanovNJP2008, QiaoPRL2009}.
This is in strong contrast to the use of linear polarization, where at that moment the carbon ion phase space already forms a continuous, straight line (fig. \ref{fig3}a). 
\begin{figure}[t!]
 \includegraphics[width=8.6cm]{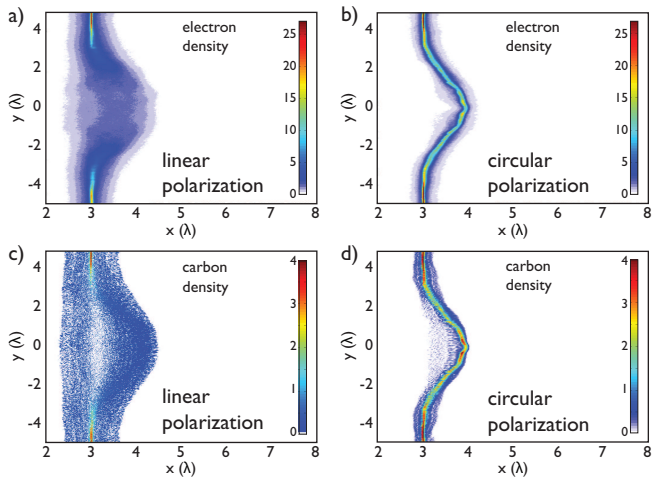}
 \caption{(color). Cycle-averaged electron (a,b) and carbon ion (c,d) density at $t=61\,$fs after the peak of the laser pulse reached the $5.3\,$nm target initially located at $x=3\lambda$. While linear polarization results in strong expansion of the target caused by hot electrons, for circularly polarized irradiation the foil is accelerated as a dense, quasi-neutral plasma bunch.
 \label{fig4}}
 \end{figure}
The striking difference in the acceleration dynamics can also be directly inferred when examining the electron and ion density distributions as observed in our simulations (see fig. \ref{fig4}). For use of circular polarization, the electron population maintains its structure as a thin layer of high density being pushed by the laser in forward direction. As a consequence, carbon ions co-propagate with the compressed electron cloud and the whole focal volume is accelerated as a quasi-neutral dense plasma bunch by the laser radiation pressure. 
Since all ion species move at equal velocity, RPA-driven protons attain only $30/12=2.5\,$MeV energy, thus not being visible in the continuous experimental spectrum reaching up to $\sim10\,$MeV (fig. \ref{fig2}b).
This scenario is contrary to the case of linear polarization (fig. \ref{fig4}a,c), where the foil electrons are heavily heated by the laser. Accordingly, electrons have already spread significantly and the electron density in the focal spot center, where electrons gain their highest energies, is considerably reduced. Although the acceleration of carbon ions is still asymmetric, favoring the laser forward direction (see also \cite{HenigPRL2009b}), it is dominated by the sheath field of the expanding electrons.\\
However, given the laser parameters used in the presented experiment, the narrow isolated quasi-monoenergetic peak in the carbon spectrum as it is present right after the end of laser-target interaction does not preserve its shape upon further propagation of the ion beam (fig. \ref{fig2}b,d). Even though the apex energy stays constant, the spectral distribution broadens and partially merges with the low energetic ion population which at this point still gains energy, resulting in the carbon ion spectrum as observed experimentally. This behavior can be attributed to the considerable deformation of the foil plasma by the tightly focussed gaussian laser spot (fig. \ref{fig4}b,d). Owing to the thus no longer normal but rather oblique incidence on the bent plasma surface, particularly at the end of the laser-target interaction, perpendicular electric field components are present. Those efficiently heat electrons located in the warped spatial regions, which then quickly spread around the target, causing the mono-energetic peak in the carbon spectrum to broaden as well as ions in the low energetic part of the distribution to gain further energy and close the gap to the RPA-generated population. We note, that this temporal evolution of the foil plasma shape was already discussed theoretically in \cite{KlimoPRSTAB2008}, where in order to prevent the distortion of the ion spectrum an upper limit for the laser pulse duration is given by $t_{max}\simeq \sqrt{2w_0 c \rho d/I_0}=176\,\text{fs}$.
Even though the laser pulse duration used in the presented experimental study is significantly shorter than $t_{max}$, a spectral deformation is still observed, taking place after the end of the laser-target interaction.\\
In summary, we have presented experimental investigations on ion acceleration from nm-thin DLC foil targets irradiated by linearly and circularly polarized, highly intense laser pulses. A strong decrease in the number of hot electrons is observed for use of circular polarization, resulting in a pronounced peak centered at 30\,MeV in the carbon C$^{6+}$ ion spectrum at the optimum foil thickness of 5.3\,nm, which is in excellent agreement with the condition $a_0\,\simeq\,\sigma$ \cite{YanPRL2008}. Compared to \cite{HegelichN2006}, we demonstrate a more than 40 times increase in conversion efficiency when considering an identical energy spread around the apex. 2D PIC-simulations have been carried out, giving clear evidence that those ions are for the first time dominantly accelerated in a phase-stable way by the laser radiation pressure. While linear polarization gives rise to the generation of a large number of highly energetic electrons, causing the target plasma to expand rapidly, in case of circular polarization electrons and ions co-propagate as a dense, quasi-neutral plasma bunch over the whole duration of the laser pulse. Being recently widely studied in theory, our comparative measurements provide the first experimental proof of the feasibility of radiation pressure acceleration to become the dominant mechanism for ion acceleration when circular polarization is used. These results are a major step towards highly energetic, mono-chromatic ion beams generated at high conversion efficiencies as demanded by many potential applications. Those include fast ignition inertial confinement fusion (ICF) as well as oncology and radiation therapy.
\begin{acknowledgments}
This work was supported by DFG through Transregio SFB TR18 and the DFG Cluster of
Excellence Munich-Centre for Advanced Photonics (MAP). A.~Henig, D.~Kiefer and D.~Jung acknowledge financial support
from IMPRS-APS, J.~Schreiber from DAAD, X.~Q.~Yan from the
Humboldt foundation and NSFC(10855001).
\end{acknowledgments}

\end{document}